\documentclass[review]{elsarticle} 

\usepackage{amssymb}
\usepackage{amsfonts, amsmath, amssymb}
\usepackage{amsthm}
\usepackage{graphicx}
\usepackage{float}
\usepackage{enumitem}

\usepackage[utf8]{inputenc} 
\usepackage{enumitem}
\usepackage{lipsum}  

\usepackage{graphicx}
\usepackage{subcaption}
\usepackage[utf8]{inputenc}
\usepackage{url}

\begin{document}

\begin{frontmatter}



\title{SEIRD model in heterogenous populations: The role of commuting and social inequalities in the COVID-19 dynamics.}


\author[Gondim]{João A. M. Gondim}
\author[Tanaka]{Thiago Yukio Tanaka}

\address[Gondim]{Unidade Acad\^{e}mica do Cabo de Santo Agostinho, Universidade Federal Rural de Pernambuco, Cabo de Santo Agostinho, PE, Brazil}
\address[Tanaka]{Departamento de Matem\'{a}tica, Universidade Federal Rural de Pernambuco, Recife, PE, Brazil}

\begin{abstract}
In this paper we analyze the effects of commuting and social inequalities for the epidemic development of the novel coronavirus (COVID-19). With this aim we consider a SEIRD (susceptible, exposed, infected, recovered and dead by disease) model without vital dynamics in a population divided into patches that have different economic resources and in which the individuals can commute from one patch to another (bilaterally). In the modeling we choose the social and commuting parameters arbitrarily. We calculate the basic reproductive number $R_0$ with the next generation approach and analyze the sensitivity of $R_0$ with respect to the parameters. 
Furthermore, we run numerical simulations considering a population divided into two patches to bring some conclusions on the number of total infected individuals and cumulative deaths for our model considering heterogeneous populations.

\end{abstract}



\begin{keyword}
COVID-19 \sep Commuting \sep SEIRD model \sep Heterogeneous populations \sep Social inequalities.


\end{keyword}

\end{frontmatter}


\section{Introduction}
\label{}

In the beginning of December 2019 a new type of coronavirus disease was identified in the city of Wuhan, the largest city of the Hubei province in China. Currently known (officially) by the name COVID-19, this novel coronavirus was only reported to the world by the end of the same month \cite{riou2020pattern}.
According to the World Health Organization, with the data published by February 21st, just three months since the first case, over 75 thousand cases were reported just in China, with a fatality ratio around $3\%$. Also, at that same time other 26 countries have confirmed infected cases \cite{boletim}.

Since then, the spread of the COVID-19 became one of the most problematic public health case of epidemic diseases around the world, causing relevant socioeconomic impacts: primarily by the relatively high death rate causing deaths by the health complications or even by the saturation of health care systems, and secondarily since this is a new type of coronavirus, so the majority of the individuals are susceptible to the disease. Hence, until the the development of the vaccine, it will stagnate the entire system of social interaction leading to the increase in unemployment rates, weakening the health of the population and many other complications. 

At the present time, early August of 2020, the World Health Organization \cite{who} reported over 18.6 million confirmed cases of infected including over 700 thousand deaths. The Americas concentrate over $50\%$ of all infected cases with over 380 thousand deaths, Europe concentrates the highest mortality rate with approximately $18,5\%$ of all infected cases with over 216 thousand deaths representing over $30\%$ of all death cases. In total, COVID-19 is affecting 213 countries and territories around the world and 2 international conveyances \cite{worldometers}.

Many efforts have been made by the scientific community around the world in an attempt to propose models that allow mapping the spread and predictions about numerical issues involving this current pandemic. Although these models do not portray reality with absolute precision, they have helped us guide control measures (such as quarantines and educational campaigns of social distance, use of masks, cleaning and hygiene tips, and many others) to minimize the effects of the pandemic with relative success. 

In \cite{mammeri2020reaction},
the author proposes a SEIR model with a partial differential reaction-diffusion system to explain the spread of COVID-19 in France and analyzed the different situations without lockdown and with partial lockdown scenarios. In \cite{chimmula2020time}, the authors developed a forecasting model of COVID-19 outbreak in Canada using time series analysis with a machine learning approach. In \cite{castilho2020assessing}, an age-structured SEIR model was studied and the efficiency of age-oriented control strategies were assessed. In \cite{gondim2020optimal}, the authors present an optimal control result for an age-structured SEIRQ model by using quarantine strategies as the control force. 

For results more connected this work, one can see \cite{belik2011natural}, where the authors investigated a model for spatial epidemics and analyzed the effects of human mobility patterns to spatial spread of an infectious disease. Also, in \cite{contreras2020multi} the authors proposed a SEIRA model for the spread of COVID-19 in a heterogeneous population. Finally, as we can see in \cite{van2020covid} and \cite{dyer2020covid},
studies have indicated that pandemic scenarios affectsdifferently Afro-Americans and immigrants in the United States due to unfavorable socioeconomic conditions. It is stated that African Americans represent three-quarters of the total deaths in U.S.

In Brazil, as in many other Latin America countries, due to the great socioeconomic inequalities and the partial lockdowns adopted with relatively low time of implementation (together with the large underreporting of cases) it is expected that the dynamics between locations with greater and lesser financial resources will also have great divergences in the epidemic scenarios since individuals with fewer resources in addition to not having large and efficient access to heath facilities, may not be able to comply with isolation measures (such as quarantines) and precautions (such as the use of masks). Inspired by the above comments, in this paper we present a SEIRD model for COVID-19 with a population divided into patches and we analyze the influence of two important factors: social inequalities and commuting between patches.

This work is organized as follows: In Section 2 we present our SEIRD model, explain the epidemic parameters and compute the basic reproductive number $R_0$ by the next generation approach. In Section 3 we fit the parameters to real data from Brazil and analyze the sensitivity of $R_0$ with respect to them. In Section 4 we present  numerical simulations considering the dynamics with a population divided into two patches and analyze the effects of commuting and social inequalities. Finally, we present the conclusion of this work in Section 5. 

\section{Model structure and basic reproductive number}
\label{}

\begin{figure}[ht]
	\centering
	\includegraphics[scale=0.4,trim={3.5cm 1.0cm 3.5cm 0.5cm}]{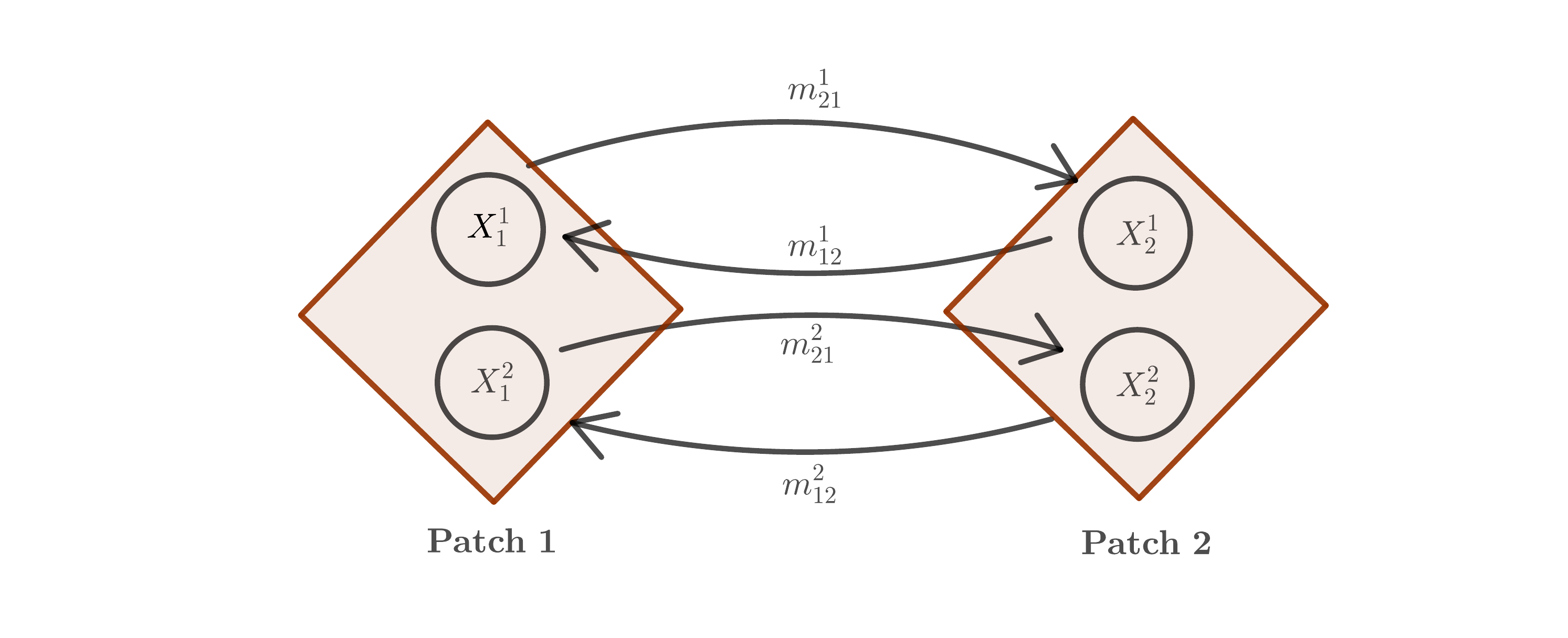}
	\caption{Diagram for commuting in a population divided into 2 patches.}
	\label{diagram-commute} 
\end{figure}

As proposed by \cite{belik2011natural}, we consider a population that is divided into $n$ patches, which could represent countries, regions, cities, or even parts of a city. People from one patch can travel to other patches, and the model distinguishes individuals by their home locations. Residents of patch $j$ that are presently in patch $i$ commute to patch $k$ at a rate $m_{ki}^j$. A diagram for the $n = 2$ case for a generic population denoted by $X$ is displayed in Figure  \ref{diagram-commute}. 

\begin{figure}[ht]
	\centering
	\includegraphics[scale=0.2,trim={3.5cm 1.0cm 3.5cm 0.5cm}]{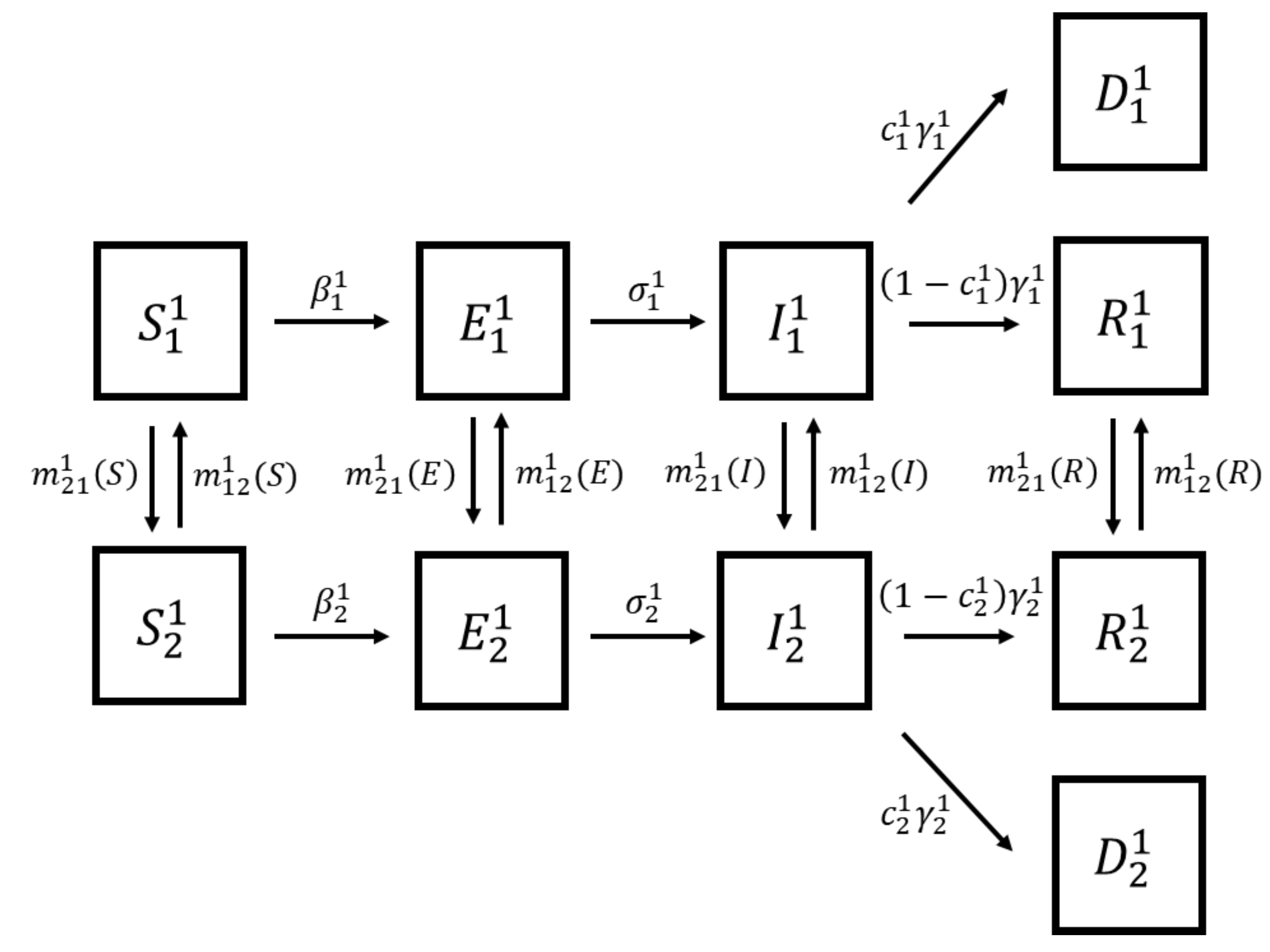}
	\vspace{0.5em}\caption{Diagram of commuting and disease progression for individuals who live in patch 1 in a population divided into 2 patches.}
	\label{diagram-seird} 
\end{figure}

The population of residents of patch $j$ which are presently in patch $i$ is also divided into four epidemiological classes. These are the susceptible, exposed, infected and recovered individuals, denoted by $S_i^j$, $E_i^j$, $I_i^j$ and $R_i^j$. We also include $D_i^j$ to represent deaths due to the disease. A diagram for the $n =2 $ case describing the commuting dynamic in one of the patches and the progression in classes is displayed in Figure \ref{diagram-seird}.

Our SEIRD model is, then, composed of $5n^2$ equations, described in \eqref{model1} for $1 \leq i,j \leq n$.

\begin{equation}
    \left\{ \begin{aligned}
    \frac{dS_i^j}{dt} &= - \beta_i^j \frac{S_i^j}{N_i} \sum_{k=1}^n I_i^k + \sum_{k=1}^n \left(m_{ik}^{j}(S)S_k^j - m_{ki}^{j}(S)S_i^j\right) \\
    \frac{dE_i^j}{dt} &= \beta_i^j \frac{S_i^j}{N_i} \sum_{k=1}^n I_i^k - \sigma_i^j E_i^j + \sum_{k=1}^n \left(m_{ik}^{j}(E)E_k^j - m_{ki}^{j}(E)E_i^j\right) \\
    \frac{dI_i^j}{dt} &= \sigma_i^j E_i^j - \gamma_i^j I_i^j + \sum_{k=1}^n \left(m_{ik}^{j}(I)I_k^j - m_{ki}^{j}(I)I_i^j\right) \\
    \frac{dR_i^j}{dt} &= \left(1-c_i^j\right)\gamma_i^j I_i^j + \sum_{k=1}^n \left(m_{ik}^{j}(R)R_k^j - m_{ki}^{j}(R)R_i^j\right) \\
    \frac{dD_i^j}{dt} &= c_i^j \gamma_i^j I_i^j \\
    \end{aligned} \right. ,
    \label{model1}
\end{equation}

In \eqref{model1}, $$N_i = \sum_{j=1}^n\left( S_i^j+E_i^j+I_i^j+R_i^j\right)$$
denotes the number of residents presently in patch $i$ from the other patches. All parameters are nonnegative. For $C \in \{S,E,I,R\}$, representing a generic epidemiological class, $m_{ki}^{j}(C)$ is the commuting rate from patch $i$ to patch $k$ for individuals of class $C$ that are residents of patch $j$. It will be assumed that $m_{ii}^{j}(C) = 0$ for all $i, j$ since there is no commuting in these cases. We denote a generic parameter $\rho_i^j$ for residents of patch $j$, presently in patch $i$. The remaining parameters are described in Table \ref{parametros}. It is assumed that the total population is constant because the analysis considers only a short time in comparison to the demographic time scale. Hence, the equations in \eqref{model1} do not include vital parameters.

\begin{table}[!h]
 	\centering
 	\caption{Description of parameters.}
 	\label{parametros}
 	\begin{tabular}{cc}
 		\hline  
 		Parameter  &   Description    \\ 
 		 \hline

 		$\beta_i^j$ & Transmission coefficient.\\
 		$\sigma_i^j$ & Exit rate of exposed class. \\
 		$\gamma_i^j$ & Exit rate of infected class. \\
 		$c_i^j$ & Case fatality ratio due to the disease. \\
 		 \hline
 	\end{tabular}%
\end{table}

We now show how to calculate the basic reproductive number, $R_0$, for model \eqref{model1}. This will be done by a next generation approach. Due to the high number of dimensions, we only consider the case $n = 2$. Firstly, we need to find the disease-free equilibrium, which is defined by $E_i^j = I_i^j = R_i^j = 0$ for all $i,j$. The equations for $\frac{dS_i^{j}}{dt}$ become the linear system
\begin{equation}
\left\{
\begin{aligned}
    \frac{dS_1^1}{dt} &= m_{12}^1(S)S_2^1 - m_{21}^1(S)S_1^1 \\
    \frac{dS_2^1}{dt} &= m_{21}^1(S)S_1^1 - m_{12}^1(S)S_2^1 \\
    \frac{dS_1^2}{dt} &= m_{12}^2(S)S_2^2 - m_{21}^2(S)S_1^2 \\
    \frac{dS_2^2}{dt} &= m_{21}^2(S)S_1^2 - m_{12}^2(S)S_2^2 \\
\end{aligned}
\right. .
    \label{eqS-dfe}
\end{equation}

Notice that the total population in each patch is constant. At the equilibrium $$S_1^1 = S_1^{1*}, \quad S_2^1 = S_2^{1*}, \quad S_1^2 = S_1^{2*}, \quad S_2^2 = S_2^{2*},$$ we have
\begin{equation}
    S_1^{1*} + S_2^{1*} = S_1^1(0) + S_2^1(0), \quad S_1^{2*} + S_2^{2*} = S_1^2(0) + S_2^2(0)
    \label{dfe-eq}
\end{equation}
and

\begin{equation}
    m_{12}^1(S)S_2^1 = m_{21}^1(S)S_1^1, \quad
    m_{12}^2(S)S_2^2 = m_{21}^2(S)S_1^2 
    \label{eqS-dfe2}
\end{equation}


Solving the system that arises from \eqref{dfe-eq} and \eqref{eqS-dfe2}, one can see that the disease-free equilibrium is 
\begin{equation}
    \begin{aligned}
        S_1^{1*} = \frac{m_{12}^1(S)\left(S_1^1(0)+S_2^1(0)\right)}{m_12^1(S)+m_21^1(S)} \quad & S_1^{2*} = \frac{m_{12}^2(S)\left(S_1^2(0)+S_2^2(0)\right)}{m_12^2(S)+m_21^2(S)} \\[4pt]
        S_2^{1*} = \frac{m_{21}^1(S)\left(S_1^1(0)+S_2^1(0)\right)}{m_12^1(S)+m_21^1(S)} \quad & S_1^{2*} = \frac{m_{21}^2(S)\left(S_1^2(0)+S_2^2(0)\right)}{m_12^2(S)+m_21^2(S)} 
    \end{aligned}.
    \label{dfe-formula}
\end{equation}

Finally, $R_0$ is given by the spectral radius of the next generation matrix (see \cite{diekmann1990definition}, \cite{martcheva2015introduction}) $K = FV^{-1}$, where

$$F = \left( \begin{array}{cccccccc} 0 & 0 & 0 & 0 & \frac{\beta_1^1 S_1^{1*}}{S_1^{1*}+S_1^{2*}} & \frac{\beta_1^1 S_1^{1*}}{S_1^{1*}+S_1^{2*}} & 0 & 0 \\ 0 & 0 & 0 & 0 & \frac{\beta_1^2 S_1^{2*}}{S_1^{1*}+S_1^{2*}} & \frac{\beta_1^2 S_1^{2*}}{S_1^{1*}+S_1^{2*}} & 0 & 0 \\ 0 & 0 & 0 & 0 & 0 & 0 & \frac{\beta_2^1 S_2^{1*}}{S_2^{1*}+S_2^{2*}} & \frac{\beta_2^1 S_2^{1*}}{S_2^{1*}+S_2^{2*}} \\ 0 & 0 & 0 & 0 & 0 & 0 & \frac{\beta_2^2 S_2^{2*}}{S_2^{1*}+S_2^{2*}} & \frac{\beta_2^2 S_2^{2*}}{S_2^{1*}+S_2^{2*}} \\ 0 & 0 & 0 & 0 & 0 & 0 & 0 & 0 \\ 0 & 0 & 0 & 0 & 0 & 0 & 0 & 0 \\ 0 & 0 & 0 & 0 & 0 & 0 & 0 & 0 \\ 0 & 0 & 0 & 0 & 0 & 0 & 0 & 0 \\ \end{array} \right)$$ 
and

{\scriptsize $$V = \left(\arraycolsep=1.4pt\def\arraystretch{2.2}
 \begin{array}{cccccccc} D_1^1 & 0 & -m_{12}^{1}(E) & 0 & 0 & 0 & 0 & 0 \\ 0 & D_1^2 & 0 & -m_{12}^{2}(E) & 0 & 0 & 0 & 0 \\ -m_{21}^{1}(E) & 0 & D_2^1 & 0 & 0 & 0 & 0 & 0 \\ 0 & -m_{21}^{2}(E) & 0 & D_2^2 & 0 & 0 & 0 & 0 \\ -\sigma_1^1 & 0 & 0 & 0 & D_3^1 & 0 & -m_{12}^{1}(I) & 0 \\ 0 & -\sigma_1^2 & 0 & 0 & 0 & D_3^2 & 0 & m_{12}^{2}(I) \\ 0 & 0 & -\sigma_2^1 & 0 & -m_{21}^{1}(I) & 0 & D_4^1 & 0 \\ 0 & 0 & 0 & -\sigma_2
^2 & 0 & -m_{21}^{2}(I) & 0 & D_4^2 \\ \end{array} \right)$$}
with diagonal elements as in Table \ref{diagonal}.

\begin{table}[!h]
 	\centering
 	\caption{Diagonal elements of $V$ ($j \in \{1,2\}$).}
 	\label{diagonal}
 	\begin{tabular}{cc}
 		\hline  
 		Element  &   Formula    \\ 
 		 \hline
 		$D_1^j$ &  $m_{21}^{j}(E)+\sigma_1^j$.\\ 
 		$D_2^j$ &  $m_{12}^{j}(E)+\sigma_2^j$. \\
 		$D_3^j$ &  $\gamma_1^j+m_{21}^{j}(I)$.\\
 		$D_4^j$ &  $\gamma_2^j+m_{12}^{j}(I)$. \\
 		 \hline
 	\end{tabular}%
\end{table}

Due to the complexity of $K$ and its eigenvalues, we do not give an explicit formula for $R_0$, which would not have much analytical use with so many parameters. Instead, we perform a numerical sensitivity analysis in the next Section.


\section{Data fitting and $R_0$ sensitivity analysis}
\label{}

In this Section we consider model \eqref{model1} with 2 patches, which represent two regions of a given large city of Brazil. The parameters for these patches will be chosen arbitrarily to model social inequalities. One patch will consist of wealthier individuals, who are better able to self isolate and have greater access to hospitals.

In order to allow us to choose the parameters better, we start by fitting the number of infected people in Brazil for the first 20 days of the outbreak in a SEIR model with only one patch. The data from \cite{worldometers} is shown in Table \ref{data}. The model is

\begin{equation}
        \left\{ \begin{aligned}
    \frac{dS}{dt} &= - \beta \frac{S I}{N} \\
    \frac{dE}{dt} &= \beta \frac{SI}{N} - \sigma E   \\
    \frac{dI}{dt} &= \sigma E - \gamma I   \\
    \frac{dR}{dt} &= \gamma I  
    \end{aligned} \right. .
    \label{model2}
\end{equation}

\begin{table}[!h]
 	\centering
 	\caption{Active number of cases in Brazil for the 20 first days of the Covid-19 outbreak.}
 	\label{data}
 	\begin{tabular}{cccccc}
 		\hline  
 		Day & Date  & Number of cases & Day & Date & Number of cases    \\ 
 		 \hline
 		1 & February 25 & 1 & 11 & March 6 & 13 \\
 		2 & February 26 & 1 & 12 & March 7 & 19 \\
 		3 & February 27 & 1 & 13 & March 8 & 25 \\
 		4 & February 28 & 1 & 14 & March 9 & 25 \\
 		5 & February 29 & 2 & 15 & March 10 & 34 \\
 		6 & March 1 & 2 & 16 & March 11 & 52 \\
 		7 & March 2 & 2 & 17 & March 12 & 77 \\
 		8 & March 3 & 2 & 18 & March 13 & 150 \\
 		9 & March 4 & 3 & 19 & March 14 & 150 \\
 		10 & March 5 & 8 & 20 & March 15 & 198 \\
 		 \hline
 	\end{tabular}%
\end{table}




Brazil's population will be rounded to 200 million people. Now, a minimization routine based on the least squares method is used to find parameters $\beta$ and $\gamma$ to the data in Table \ref{data}. Since the latency period is estimated to be 5.2 days \cite{li2020early}, we assume that $\sigma = 1/5.2$. The minimization algorithm is an adaptation of one that is available in \cite{martcheva2015introduction}, starting with an initial guess of $\beta = 1$ and $\gamma = 0.1$ and initial conditions $S(0) = 200.000.000$, $E(0) = 0$, $I(0) = 1$ and $R(0) = 0$. The results are 
\begin{equation}
    \beta^* = 0.9230, \quad \gamma^* = 0.0458.
    \label{fittedparameters}
\end{equation}

\begin{figure}[ht]
	\centering
	\includegraphics[scale=0.5,trim={3.5cm 9cm 3.5cm 9.5cm}]{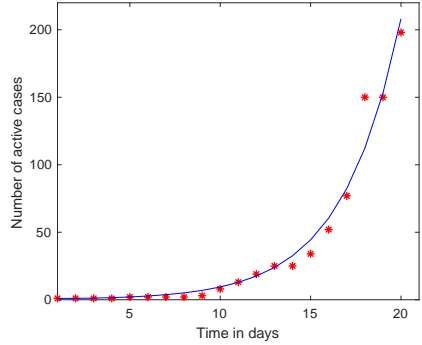}
	\caption{Data from Table \ref{data} along with the infected curve for model \eqref{model2}. Parameters are $\beta = 0.92317$, $\sigma = 1/5.2$ and $\gamma = 0.04582$.}
	\label{fig1} 
\end{figure}

The data from Table \ref{data} and the infected curve with the parameters chosen as above are shown in Figure \ref{fig1}.

We assume that patch 1 is the wealthier one. For $C \in \{S,E,R\}$, we assume that 
\begin{equation}
    m_{ij}^{k}(C) = m_{ij}^{k}, \quad i,j,k \in \{1,2\},
\end{equation}
i.e., the commuting rates do not depend on the epidemiological class for non-infected individuals. In their case, we suppose that the symptoms reduce their mobility, so we take
\begin{equation}
    m_{ij}^{k}(I) = \frac{m_{ij}^k}{2}, \quad i,j,k \in \{1,2\}.
\end{equation}

Furthermore, we assume that $m_{ji}^i << m_{ij}^i$, $i, j \in \{1,2\}$, that is, individuals spend much more time in their base locations, and that $m_{21}^1 << m_{12}^2$, i.e., residents of patch 2 are more likely to go to patch 1 than the other way around. In a specific case, this can be seen as follows: patch 1 concentrates more places of entertainment and commerce compared to patch 2. For our numerical analysis of $R_0$ and the simulations in the next Section, we take 
\begin{equation}
    m_{21}^1 = 1/9, \quad m_{12}^1 = 1, \quad m_{12}^2 = 1.5, \quad m_{21}^2 = 3.
\end{equation}

We also assume that the number of residents in each patch is 30000. From \eqref{dfe-formula}, the disease-free equilibrium is, then,
\begin{equation}
    S_1^{1*} = 27000, \quad S_2^{1*} = 3000, \quad S_1^{2*} = 10000, \quad S_2^{2*} = 20000.
    \label{dfe-values}
\end{equation}

For the epidemiological parameters, we suppose that $\beta_1^1 = \beta_1^2$, $\beta_2^1 = \beta_2^2$, $\sigma_{ij}^k = \sigma^*$ for $i,j,k\in\{1,2\}$, $\gamma_1^1 = \gamma_2^1$ and $\gamma_1^2 = \gamma_2^2$. This means that the transmission coefficients will be considered as a property of the patch in which individuals currently are, whereas the recovery rates depend only on the patch they live in.

Our choices, then, will be $\beta_1^1 = \beta_1^2 = \beta^*$, $\gamma_1^1 = \gamma_2^1 = \gamma^*$ and $\beta_2^1 = \beta_2^2 = (1+p)\beta^*$, $\gamma_1^2 = \gamma_2^2 = \gamma^*/(1+p)$, where $p>0$ is a parameter that reflects the social inequalities between the two patches. We consider $p = 0.5$. A summary of parameter values is available in Table \ref{resumo}.

\begin{table}[!h]
 	\centering
 	\caption{Summary of parameter values for the sensitivity analysis.}
 	\label{resumo}
 	\begin{tabular}{cccc}
 		\hline  
 		Parameter & Value  & Parameter & Value   \\ 
 		 \hline
 		$m_{12}^1$ & $1$ & $m_{21}^1$ & $1/9$ \\
 		$m_{12}^2$ & $1.5$ & $m_{21}^2$ & $3$ \\
 		$\beta^*$ & $0.9230$ & $\gamma^*$ & $0.0458$ \\
 		$\sigma$ & $1/5.2$ & $p$ & $0.5$ \\
 		$S_1^{1*}$ & $27000$ & $S_2^{1*}$ & $3000$ \\
 		$S_1^{2*}$ & $10000$ & $S_2^{2*}$ & $20000$ \\
 		 \hline
 	\end{tabular}%
\end{table}




The normalized forward sensitivity index of $R_0$ (see \cite{rosa2019optimal,chitnis2008determining}) is given by 
\begin{equation}
    \Upsilon_\rho^{R_0} = \frac{\partial R_0}{\partial \rho}\cdot \frac{\rho}{R_0}, 
    \label{index}
\end{equation}
where $\rho$ is a parameter. This number gives the percentage change in $R_0$ with respect to a percentage change in $\rho$ (see \cite{martcheva2015introduction}). For example, if $\Upsilon_\rho^{R_0} = 0.1$, then a $1\%$ increase in $\rho$ increases $R_0$ in $0.1\%$. After computing the partial derivatives numerically, we find the results displayed in Table \ref{sensitivity}.

\begin{table}[!h]
 	\centering
 	\caption{Sensitivity of $R_0$ with respect to the parameters.}
 	\label{sensitivity}
 	\begin{tabular}{cccc}
 		\hline  
 		Parameter & Sensitivity index  & Parameter & Sensitivity index   \\ 
 		 \hline
 		$m_{12}^1$ & $0.0215$ & $m_{21}^1$ & $-0.0221$ \\
 		$m_{12}^2$ & $-0.1142$ & $m_{21}^2$ & $0.1141$ \\
 		$\beta^*$ & $1.0000$ & $\gamma^*$ & $-0.9898$ \\
 		$\sigma$ & $0.0003$ & $p$ & $0.5159$ \\
 		 \hline
 	\end{tabular}%
\end{table}

Notice that the effect of changes in the commuting rates on $R_0$ is small in comparison to other parameters. This shows that, even though travel restrictions are useful methods of controlling the spread of as epidemic across patches, they are not as effective as measures that affect the transmission coefficient (such as wearing masks and social distancing) and the recovery rate (such as screening measures and the isolation of infected individuals) \cite{castilho2006optimal}.

However, the rates corresponding to the mobility of residents of the poorer patch induce percentage changes in $R_0$ around five times those corresponding to the wealthier patch. Moreover, we see that leaving your base patch decreases $R_0$, whereas returning home increases it.

On the other hand, we see that a $10\%$ reduction in $p$ produces a $5.1\%$ reduction in $R_0$, so reducing social inequalities is paramount in regard to the control of future epidemic outbreaks.

\section{Numerical simulations}
\label{}

In this Section we use the parameters from Section 3 to gather information about the disease spread in a population divided into two patches. We aim to answer the following questions:

\begin{enumerate}
    \item[(i)] How does the maximum of the infected curve depend on the parameter $p$?
    \item[(ii)] How many deaths are due to traveling between patches?
    \item[(iii)] How are these extra deaths distributed in the populations of the two patches?
\end{enumerate}

As initial conditions, we consider susceptible populations as the equilibrium values from Table \ref{resumo}, zero exposed, recovered and deaths, and one infected individual, which is a resident of the wealthier patch. This happened in a few large cities of Brazil such as São Paulo \cite{correio} and Recife \cite{folha}, where the first cases were infected after trips to Europe.

\begin{figure}[h!]
	\centering
	\includegraphics[scale=0.75]{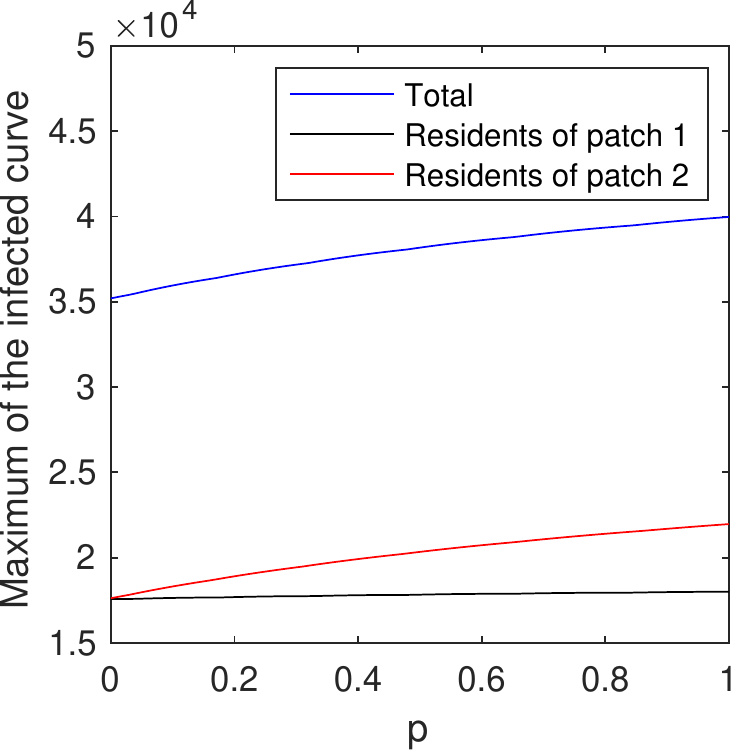}
	\caption{The maximum of the infected curve as a function of $p$.}
	\label{fig2} 
\end{figure}

To analyze the first question, we plot, in Figure \ref{fig2}, the maximum of the total infected curve $$I_1^1(t)+I_2^1(t)+I_1^2(t)+I_2^2(t)$$ as a function of $p \in [0,1]$. Notice that, as $p$ increases, the maximum also increases. However, if we do the same for $$I_1^1(t) + I_2^1(t)$$ and $$I_1^2(t)+I_2^2(t),$$ we see that the maximum is essentially constant with regard to $p$ for residents of the wealthier patch, so the increment in the overall maximum comes at the expense of the poorer patch.

A bigger number of simultaneously infected individuals poses a serious problem, because it could lead to the saturation of the healthcare system \cite{learning}. Figure \ref{fig2} suggests also that the more unequal societies are, the bigger the toll on hospitals and other facilities is on the poorer regions, whose residents already have less access to it to begin with.

\begin{figure}[h!]
	\centering
	\includegraphics[scale=0.68,trim={3.5cm 10.5cm 3.5cm 10.5cm}]{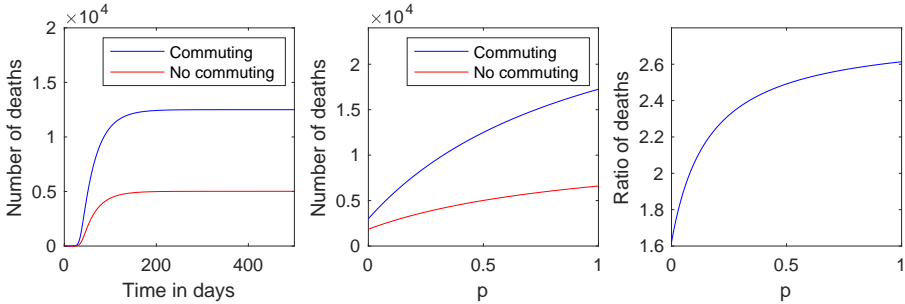}
	\caption{On the left, plots of the number of cumulative number of deaths as a function of time for $p = 0.5$ and scenarios with and without commuting. In the middle, the number of deaths is plotted as a function of $p$. On the left, the ratio of deaths with and without commuting as a function of $p$.}
	\label{fig3} 
\end{figure}

The second question is answered with Figure \ref{fig3}. On the left, it shows plots of $$D_1^1(t) + D_2^1(t) + D_1^2(t) + D_2^2(t)$$ for $p = 0.5$. The blue curve is the solution that arises from the parameters from Table \ref{resumo}, whereas the red curve represents a situation with no commuting, i.e., $m_{12}^1 = m_{21}^1 = m_{12}^2 = m_{21}^2 = 0$. 

This scenario represents a situation in which the borders of each patch are closed, for example, in an attempt to stop the spread of the disease. As such, there are people outside of their base places when the travels are stopped, which explains why, in the following simulations, there are cases and deaths of patch 2 residents even though there were no infected residents from this patch in the initial conditions.

The case fatality ratios considered in patch 1 were $c_1^1 = c_2^1 = c_0 = 0.05$, that is, 5$\%$ of infected residents from this patch die from the disease. In patch 2, we assume that the case fatalities increase with $p$ in such a way that, when $p = 0$, we have $c_1^2 = c_2^2 = c_0$, and when $p \rightarrow \infty$, we have $c_1^2 = c_2^2 = 1$. Hence, we take
\begin{equation}
    c_1^2(p) = c_2^2(p) = \frac{p+c_0}{p+1}.
    \label{cfr-p}
\end{equation}

The plot on the left of Figure \ref{fig3} shows that commuting increased the total number of deaths, given by the equilibrium values of the two curves, in around $2.5$ times. In order to determine if this was a property of the specific value of $p$ that was chosen, the other two plots of Figure \ref{fig3} exhibit the total number of deaths as a function of $p \in [0,1]$, with and without commuting, and the ratio of the numbers of deaths with commuting to without commuting. The resulting curves indeed reveal that the travels amplify the total number of deaths as social discrepancies rise.

\begin{figure}[ht]
	\centering
	\includegraphics[scale=0.75,trim={3.5cm 7.0cm 3.5cm 6.5cm}]{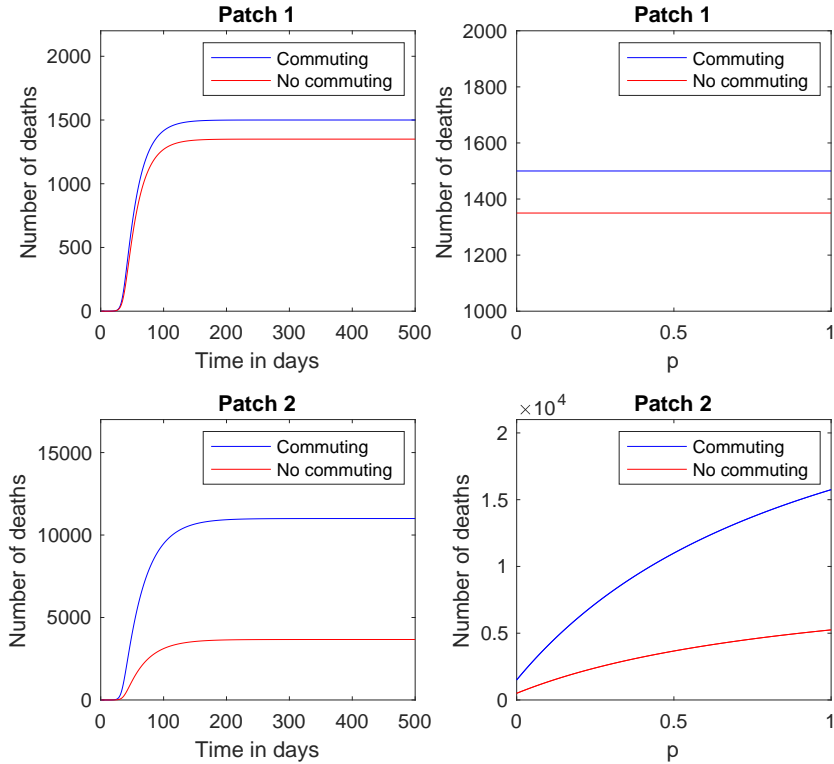}
	\caption{Plots of the number of cumulative number of deaths as a function of time for $p = 0.5$ (on the left) and of the final number of deaths as a function of $p$ (on the right). Scenarios with and without commuting are considered for residents of both patches.}
	\label{fig4} 
\end{figure}

This leads us to Figure \ref{fig4} and the answer to our last question. The plots display the cumulative number of deaths of residents from each patch as functions of time and $p$. Notice the discrepancy in how the travels increases the deaths in both patches. Moreover, the deaths do not change as $p$ varies in patch 1, so the extra deaths we saw on Figure \ref{fig3} come at the expense of patch 2.

In patch 1, whose residents are richer, the final number of deaths was multiplied by a factor of $1.11$. On the other hand, this factor was $3$ in patch 2 so, once again, the disease has much worse consequences for poorer individuals. Finally, we look at the percentages of deaths coming from each patch. For $p = 0.5$, the results from Figures \ref{fig3} and \ref{fig4} are presented in Table \ref{percentages}.

\begin{table}[!h]
 	\centering
 	\caption{Percentage of deaths in each patch for $p = 0.5$.}
 	\label{percentages}
 	\begin{tabular}{cccc}
 		\hline  
 		  &   Total & Patch 1 & Patch 2    \\ 
 		 \hline

 		With commuting & 12500 & 1500 & 11000 \\
 		Percentages & & 12\% & 88\% \\
 		Without commuting & 5016 & 1350 & 3666 \\
 		Percentages & & 27\% & 73\% \\
 		 \hline
 	\end{tabular}%
\end{table}

Once again, the effects of travel among patches are felt in a much harsher way in the poorer patch.


\section{Conclusion}
\label{}

In this paper we developed a SEIRD model for the COVID-19 epidemic in a population distributed in different patches. We assumed that individuals can travel between patches and also that the patches have different socioeconomic resources and studied the effects of commuting and social inequalities on this dynamic. The residents of each patch were divided in four epidemiological classes of a SEIR model without vital dynamics and we included a D class to represent deaths from the disease. 

We calculated the basic reproductive number using a next generation approach. 
To fit the parameters, we used data of the infected people of the first 20 days of the pandemic outbreak in Brazil in a SEIR model with a single patch and a minimization routine based on the least squares method and then we analyzed the sensitivity of the $R_0$ with respect of these parameters. 

This analysis pointed to many interesting facts. The mobility rates of residents in the poorer patch causes percentages variations in $R_0$ around five times higher when compared to the wealthier one. Also, the travels provoke different variations on $R_0$: leaving your base patch reduces it whereas returning increases it. Another result is that the mobility rates between patches cause only small variations in $R_0$, thus indicating that avoiding commuting is not as effective a strategy as measures that directly affect the infection or recovery rates. For $p$, the parameter reflecting social inequality, the analysis is quite different. We observed that reductions of $10\%$ in $p$ causes reductions around $5\%$ in the basic reproductive number, indicating that the reduction of socioeconomic inequalities is an effective strategy in the control of epidemic outbreaks.

We did numerical simulations of the proposed model considering two patches and supposing that one is wealthier then the other. These results (present in Section 5) lead us to some important conclusions.  The first one is that social inequalities play an important role in the concentration of infected individuals. To be more precise, the maximum of the infected curve increases as the parameter $p$ increases. On the other hand, the total cases of the wealthier patch remain almost constant with the variance of $p$, therefore the most affected patch is the poorer one. 

The second is that mobility between patches directly contributes to the increase of the cumulative number of deaths. It is important to observe that in this situation there is a vast difference in the death rate when we compare commuting and no commuting scenarios in each patch: in the wealthier one we have a difference around $11\%$, whereas the approximated difference is of $200\%$ in the poorer one.

The percentages of deaths coming from the poorer patch decreased in $15\%$ in a scenario with no mobility between patches, for $p = 0.5$. Therefore we conclude that both human mobility and social inequalities represent important facts to be considered in future models. 

We expect that our model with the results presented in this paper help the scientific community to a better understanding on how human mobility and social inequalities affect the evolution of the COVID-19 pandemic and similar structures, which could lead the authorities to better public-health policies to control or minimize the effects of epidemics. Finally, it is important to mention that one can adapt our model to consider important others dynamical processes in Epidemiology, Ecology or Biology.




 \bibliographystyle{elsarticle-num} 
 \bibliography{references.bib}

\begin{thebibliography}{10}
\expandafter\ifx\csname url\endcsname\relax
  \def\url#1{\texttt{#1}}\fi
\expandafter\ifx\csname urlprefix\endcsname\relax\def\urlprefix{URL }\fi
\expandafter\ifx\csname href\endcsname\relax
  \def\href#1#2{#2} \def\path#1{#1}\fi

\bibitem{riou2020pattern}
J.~Riou, C.~L. Althaus, Pattern of early human-to-human transmission of wuhan
  2019 novel coronavirus (2019-ncov), december 2019 to january 2020,
  Eurosurveillance 25~(4) (2020) 2000058.

\bibitem{boletim}
Boletim epidemiológico 03 - doença pelo novo coronavírus - covid-19,
  \url{https://www.saude.gov.br/images/pdf/2020/fevereiro/21/2020-02-21-Boletim-Epidemiologico03.pdf},
  accessed: 2020-08-07.

\bibitem{who}
World health organization,
  \url{https://www.who.int/emergencies/diseases/novel-coronavirus-2019},
  accessed: 2020-08-09.

\bibitem{worldometers}
Worldometers, \url{https://www.worldometers.info/coronavirus/}, accessed:
  2020-08-07.

\bibitem{mammeri2020reaction}
Y.~Mammeri, A reaction-diffusion system to better comprehend the unlockdown:
  Application of seir-type model with diffusion to the spatial spread of
  covid-19 in france, arXiv preprint arXiv:2005.03499 (2020).

\bibitem{chimmula2020time}
V.~K.~R. Chimmula, L.~Zhang, Time series forecasting of covid-19 transmission
  in canada using lstm networks, Chaos, Solitons \& Fractals (2020) 109864.

\bibitem{castilho2020assessing}
C.~Castilho, J.~A. Gondim, M.~Marchesin, M.~Sabeti, Assessing the efficiency of
  different control strategies for the covid-19 epidemic, Electronic Journal of
  Differential Equations 2020~(64) (2020) 1--17.

\bibitem{gondim2020optimal}
J.~A. Gondim, L.~Machado, Optimal quarantine strategies for the covid-19
  pandemic in a population with a discrete age structure, arXiv preprint
  arXiv:2005.09786 (2020).

\bibitem{belik2011natural}
V.~Belik, T.~Geisel, D.~Brockmann, Natural human mobility patterns and spatial
  spread of infectious diseases, Physical Review X 1~(1) (2011) 011001.

\bibitem{contreras2020multi}
S.~Contreras, H.~A. Villavicencio, D.~Medina-Ortiz, J.~P. Biron-Lattes,
  {\'A}.~Olivera-Nappa, A multi-group seira model for the spread of covid-19
  among heterogeneous populations, Chaos, Solitons \& Fractals (2020) 109925.

\bibitem{van2020covid}
A.~van Dorn, R.~E. Cooney, M.~L. Sabin, Covid-19 exacerbating inequalities in
  the us, Lancet (London, England) 395~(10232) (2020) 1243.

\bibitem{dyer2020covid}
O.~Dyer, Covid-19: Black people and other minorities are hardest hit in us
  (2020).

\bibitem{diekmann1990definition}
O.~Diekmann, J.~A.~P. Heesterbeek, J.~A. Metz, On the definition and the
  computation of the basic reproduction ratio r 0 in models for infectious
  diseases in heterogeneous populations, Journal of mathematical biology 28~(4)
  (1990) 365--382.

\bibitem{martcheva2015introduction}
M.~Martcheva, An introduction to mathematical epidemiology, Vol.~61, Springer,
  2015.

\bibitem{li2020early}
Q.~Li, X.~Guan, P.~Wu, X.~Wang, L.~Zhou, Y.~Tong, R.~Ren, K.~S. Leung, E.~H.
  Lau, J.~Y. Wong, et~al., Early transmission dynamics in wuhan, china, of
  novel coronavirus--infected pneumonia, New England Journal of Medicine
  (2020).

\bibitem{rosa2019optimal}
S.~Rosa, D.~F. Torres, Optimal control and sensitivity analysis of a fractional
  order tb model, Statistics, Optimization \& Information Computing 7~(3)
  (2019) 617--625.

\bibitem{chitnis2008determining}
N.~Chitnis, J.~M. Hyman, J.~M. Cushing, Determining important parameters in the
  spread of malaria through the sensitivity analysis of a mathematical model,
  Bulletin of mathematical biology 70~(5) (2008) 1272.

\bibitem{castilho2006optimal}
C.~Castilho, Optimal control of an epidemic through educational campaigns.,
  Electronic Journal of Differential Equations (EJDE)[electronic only] 2006
  (2006) Paper--No.

\bibitem{correio}
Primeiro caso de covid-19 no brasil completa 150 dias,
  \url{https://tinyurl.com/correio-primeiro-caso}, accessed: 2020-08-07.

\bibitem{folha}
Casal do recife são primeiros casos de coronavírus em pernambuco,
  \url{https://tinyurl.com/folha-pe}, accessed: 2020-08-07.

\bibitem{learning}
Inequality in healthcare: saturation of hospitals and death by covid-19,
  \url{https://tinyurl.com/inequality-healthcare}, accessed: 2020-08-06.

\end{thebibliography}





\end{document}